\documentclass[copyright,creativecommons]{eptcs}
\providecommand{\event}{ACT 2021}

\usepackage{amsmath}
\usepackage{amsfonts}
\usepackage{amssymb}
\usepackage{amsthm}
\usepackage{booktabs}
\usepackage{bbm}
\usepackage{tikz}
\usetikzlibrary{cd}
\usepackage{ebproof}
\usepackage{stmaryrd}
\usepackage{cmll} 

\newtheorem{definition}{Definition}
\newtheorem{theorem}[definition]{Theorem}
\newtheorem{proposition}[definition]{Proposition}

\newtheorem{example}[definition]{Example}

\title{Grounding Game Semantics in Categorical Algebra} 
\author{J\'er\'emie Koenig
  \institute{Yale University, USA}
  \email{jeremie.koenig@yale.edu}
}
\newcommand{\titlerunning}{Grounding Game Semantics in Categorical Algebra}
\newcommand{\authorrunning}{J\'er\'emie Koenig}
\newcommand\ldot{.}

\begin{document}


\maketitle

\begin{abstract} 
I present a formal connection between
\emph{algebraic effects} and \emph{game semantics},
two important lines of work in
programming language semantics
with applications in compositional
software verification.

Specifically,
the algebraic signature
enumerating the possible side-effects of a computation
can be read as a game,
and strategies for this game
constitute the free algebra for the signature
in a category of complete partial orders (\emph{cpos}).
Hence,
strategies provide a convenient model
of computations with uninterpreted side-effects.
In particular,
the operational flavor of game semantics
carries over to the algebraic context,
in the form of
the coincidence between the initial algebras and the terminal coalgebras
of cpo endofunctors.

Conversely,
the algebraic point of view
sheds new light on the strategy constructions
underlying game semantics.
Strategy models can be reformulated as
ideal completions of partial strategy trees
(free dcpos on the term algebra).
Extending the framework to multi-sorted signatures
would make this construction available
for a large class of games.
\end{abstract}

\section{Introduction} 

Writing bug-free software is notoriously hard.
Current practice encourages
comprehensive testing,
but while testing can reveal bugs
it can never completely guarantee their absence.
Therefore, for critical systems,
\emph{verification} has become the gold standard:
the desired behavior is described
as a mathematical specification,
against which the implemented system is formally proven correct
\cite{shao10}.

Over the past decade,
researchers have been able to apply this methodology
to larger and larger systems:
there are now verified compilers \cite{compcert,vellvm,cakeml},
operating system kernels \cite{sel4,popl15,osdi16},
and even verified processor designs \cite{safe,kami}.
As a result,
the construction of large-scale, heterogeneous computer systems
which are fully verified
is now within reach \cite{deepspec}.
A system of this kind would be
described end-to-end by a mathematical model,
and certified correct by a computer-checked proof,
providing a strong guarantee that
a given combination of hardware and software components
behaves as expected.

Unfortunately,
composing certified components into certified systems
is difficult.
For verification to be tractable,
the models and techniques used must often be tailored
to the component at hand.
As a result,
given two certified components developed independently,
it is often challenging to
interface their proofs of correctness
to construct a larger proof encompassing them both.
To facilitate this process,
a key task will be to establish a \emph{hierarchy}
of common models.
Using this hierarchy,
individual certified components
could continue to use specialized models,
but these models could then be embedded into more general ones,
where components and proofs of different kinds would be made interoperable.

Category theory is an important tool for this task.
It can help us characterize existing models
and compare them in a common framework.
As a systematic study of compositional structures,
it can then guide the design of more general models
capable of describing heterogeneous systems.
This paper
proposes to use this methodology to
explore connections between
two related but distinct lines of work:
\begin{itemize}
  \item
    \emph{Algebraic effects} \cite{effadq,eff}
    offer a computational reading
    of basic concepts in categorical algebra
    for the purpose of modeling, combining, and reasoning about
    side-effects in computations.
    They are a principled solution
    grounded in well-established mathematics,
    and have prompted novel and promising
    approaches to programming language design.
  \item
    \emph{Game semantics} \cite{cspgs,gamesem99}
    describe interfaces of program components
    as \emph{games} played between a component and its environment,
    characterizing the component's behavior as a \emph{strategy}
    in this game.
    This approach has been used to give compositional semantics
    to existing language features which had previously resisted
    a satisfactory treatment.
\end{itemize}
%
%
The theory of algebraic effects
is outlined in \autoref{sec:mainideas}.
After introducing game semantics,
\autoref{sec:effstrat} uses the associated techniques
to construct a \emph{strategy} model of uninterpreted algebraic effects.
This model can be characterized as an initial algebra
in a particular category of complete partial orders,
and reformulated as a completion of the term algebra.
Section \ref{sec:multisort}
proposes to extend this construction to a larger class of games
by considering different completions and multi-sorted effect signatures.

I will use the notations
$\mathbbm{1} := \{ * \}$ and
$\mathbbm{2} := \{ \mathsf{tt}, \mathsf{ff} \}$.
The set of finite sequences
over an alphabet $\Sigma$ is written $\Sigma^*$,
with $\epsilon$ as the empty sequence
and $s \cdot t$ as the concatenation of
the sequences $s$ and $t$.
With that said,
since the mathematics presented here
are ultimately intended to be mechanized in a proof assistant,
I will often prefer the use of inductive grammars
rather than sets of sequences.


\section{Models of computational side-effects} \label{sec:mainideas} 


Modeling the \emph{side-effects} of computer programs
is a long-standing research topic in programming language semantics.
I begin this paper by summarizing the underlying issues
and present the approach known as \emph{algebraic effects} \cite{effadq}.



\subsection{Monadic effects} \label{sec:monadeff} 
Programs which perform pure calculations
are straightforward to interpret mathematically.
For example,%
\begin{equation} \label{eqn:functional}
  \mathrm{abs}(x) \,:=\,
    \textbf{if } x > 0
    \textbf{ then } \textbf{return } x
    \textbf{ else } \textbf{return } {-x}
\end{equation}
can be characterized using the function
$f : \mathbb{R} \rightarrow \mathbb{R}$
which maps $x$ to $|x|$.
By contrast,
consider the program%
\begin{equation} \label{eqn:imperative}
  \mathrm{greeting}(*) \,:=\,
  (\textbf{if } \mathsf{readbit}
   \textbf{ then } \mathsf{print} \text{ ``Hi''}
   \textbf{ else } \mathsf{print} \text{ ``Hello''})
  \mathbin{;}\,
  \mathsf{stop}
\end{equation}
which reads a single bit of input,
outputs ``Hi'' or ``Hello''
depending on the value of that bit,
then terminates without producing a value.
The \emph{side-effects} performed
by the operations
$\mathsf{readbit}$, $\mathsf{print}$ and $\mathsf{stop}$
are more difficult to model.
Certainly,
(\ref{eqn:imperative}) cannot be described
as a function $g : \mathbbm{1} \rightarrow \varnothing$.

The traditional way to address this issue
is to capture the available side-effects in a \emph{monad}
$\langle T, \eta, \mu \rangle$ \cite{monads}.
Then $TX$ represents computations with a result in $X$
which may also perform side-effects.
The monad's unit $\eta : X \rightarrow TX$
corresponds to $\mathbf{return}$,
a pure computation which terminates immediately.
The multiplication $\mu : TTX \rightarrow TX$
first performs the effects of the outer computation,
then those of the computation it evaluates to.
This allows us to compose
the computations $f : A \rightarrow TB$ and $g : B \rightarrow TC$
sequentially ($;$) by
using their Kleisli composition $\mu \circ Tg \circ f$.

\begin{example} \label{ex:monadic}
To assign a meaning to the program (\ref{eqn:imperative}),
we can use the following monad in $\mathbf{Set}$:
\[
  T X := (\Sigma^* \times X_\bot)^\mathbbm{2}
  \qquad
  \eta(x) := b \mapsto (\epsilon, x)
  \qquad
  \mu \big( i \mapsto \big( s_i, j \mapsto (s'_{ij}, x_{ij}) \big) \big) :=
    b \mapsto \big( s_b \cdot s'_{bb}, \: x_{bb} \big)
\]
An element of $TX$ is a function which takes as input
the bit to be read by $\mathsf{readbit}$.
In addition to the computation's result,
which can be $\bot$ as well as a result in $X$,
the function produces a sequence of characters
from a fixed alphabet $\Sigma$.
The operations
$\mathsf{readbit}$,
$\mathsf{print}$ and
$\mathsf{stop}$
can be interpreted as:
\[
  \begin{array}{c@{\qquad}c@{\qquad}c}
    \mathsf{readbit} \in T \mathbbm{2} &
    \mathsf{print} : \Sigma^* \rightarrow T \mathbbm{1} &
    \mathsf{stop} \in T \varnothing
    \\
    \mathsf{readbit} := b \mapsto (\epsilon, b) &
    \mathsf{print}(s) := b \mapsto (s, *) &
    \mathsf{stop} := b \mapsto (\epsilon, \bot)
  \end{array}
\]
Then, the program (\ref{eqn:imperative})
can be characterized using the function
$g : \mathbbm{1} \rightarrow T\varnothing$
defined by:
\[
  g(*) := b \mapsto
  \begin{cases}
    (\text{\rm ``Hi''}, \bot) & \text{if } b = \mathsf{tt} \\
    (\text{\rm ``Hello''}, \bot) & \text{if } b = \mathsf{ff}
  \end{cases}
\]
\end{example}


\subsection{Algebraic effects} 

A long-standing issue with the monadic approach
to computational side-effects
is that in general, monads do not compose.
This makes it difficult to combine programs
which use different kinds of side-effects.
This can be addressed by
restricting our attention to monads
describing \emph{algebraic} effects.

Computations with side-effects are then seen
as \emph{terms} in an algebra.
Function symbols correspond to the available effects.
Their arities correspond to the number of possible outcomes
of the effect, 
and each argument
specifies how the computation will continue
should the corresponding outcome occur.

\begin{example} \label{ex:algebraic}
To interpret our running example,
the algebraic signature
must contain the function symbols
$\mathsf{readbit} : 2$,
$\mathsf{print}[s] : 1$ and
$\mathsf{stop} : 0$.
The behavior of the program (\ref{eqn:imperative})
can then be represented as the term:
\[
  \mathsf{readbit}\big(
    \mathsf{print}[\text{\rm ``Hi''}](\mathsf{stop}), \:
    \mathsf{print}[\text{\rm ``Hello''}](\mathsf{stop})
  \big)
\]
and visualized as the tree:
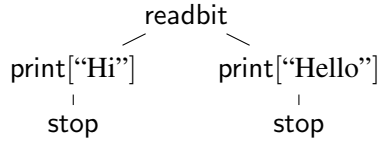
\[
  \begin{tikzpicture}[xscale=2,yscale=0.5]
    \node {$\mathsf{readbit}$}
      child {
        node {$\mathsf{print}[\text{\rm ``Hi''}]$}
        child {node {$\mathsf{stop}$}}
      }
      child {
        node {$\mathsf{print}[\text{\rm ``Hello''}]$}
        child {node {$\mathsf{stop}$}}
      };
  \end{tikzpicture}
\]
Note that $\mathsf{print}[s]$ corresponds to a \emph{family}
of operations indexed by a parameter $s \in \Sigma^*$.
\end{example}

A major advantage of this approach
is that the basic framework of universal algebra
can immediately be brought to bear.
For example,
equational theories including statements such as:
\[
  \mathsf{print}[s](\mathsf{print}[s'](x)) =
  \mathsf{print}[s \cdot s'](x)
\]
can be used to characterize the behavior of the different effects
and their possible interactions.
Algebraic theories can be combined in various ways \cite{effcomb},
making possible a compositional treatment of effects.
Below, I present a simple version of the approach,
starting with the following notion of effect signature.

\begin{definition}
An \emph{effect signature}
is a set $E$ of function symbols
together with a mapping $\mathsf{ar} : E \rightarrow \mathbf{Set}$
which assigns to each function symbol $m \in E$
an \emph{arity set} $\mathsf{ar}(m)$.
I will use the notation
\[
  E = \{ m_1 \mathbin: N_1, \: m_2 \mathbin: N_2, \: \ldots \}
\]
where $N_i = \mathsf{ar}(m_i)$ is the arity set assigned to
the function symbol $m_i$.
\end{definition}

The use of arity \emph{sets} allow us to encode
effects such as $\mathsf{readnat} : \mathbb{N}$
which have an infinite number of possible outcomes.
In this case, the argument tuples will be
families indexed by $\mathbb{N}$
and the corresponding terms
will be written as $\mathsf{readnat}(x_n)_{n \in \mathbb{N}}$.


\subsection{Initial algebras} \label{sec:initalg} 

To give a categorical account of the algebras
generated by an effect signature $E$,
we start by interpreting the signature as
an endofunctor on $\mathbf{Set}$.

\begin{definition} \label{def:sigendo}
An effect signature $E$ defines
an endofunctor $E : \mathbf{Set} \rightarrow \mathbf{Set}$
of the same name, as:
\[
  E X := \sum_{m \in E} \prod_{n \in \mathsf{ar}(m)} X
\]
\end{definition}

\noindent
The elements of $E X$
are terms of depth one with variables in $X$.
This is emphasized by the following notation
(I use underlining to distinguish term constructors
from the corresponding elements of $E$):
\[
  t \in E X ::= \underline{m} \langle x_n \rangle_{n \in \mathsf{ar}(m)}
  \qquad
  (m \in E, \: x \in X)
\]
Terms of a fixed depth $k$ can be obtained by iterating
the endofunctor as $E^k X$.
More generally,
the set of all finite terms over the signature
can be defined as follows.

\begin{definition}
Finite terms
over an effect signature $E$
with variables in $X$
are generated by the grammar:
\[
  t \in E^* X ::=
    \underline{x}
    \mid
    \underline{m}(t_n)_{n \in \mathsf{ar}(m)}
  \qquad
  (m \in E, \: x \in X)
\]
\end{definition}

\noindent
Note the use of angle brackets $\langle {-} \rangle$ for simple applications
\emph{vs.}\ parentheses $({-})$ for recursive terms.

Interpretations of the signature $E$ in a carrier set $A$
are algebras $\alpha : EA \rightarrow A$
for the endofunctor $E$.
They can be decomposed into the cotuple $\alpha = [\alpha^m]_{m \in E}$
where $\alpha^m : A^{\mathsf{ar}(m)} \rightarrow A$.
Algebras for $E$ constitute a category $\mathbf{Set}^E$
where the morphisms of type
$\langle A, \alpha \rangle \rightarrow \langle B, \beta \rangle$
are the functions $f : A \rightarrow B$
satisfying:
\[
  \begin{tikzcd}
    E A \ar[r, "\alpha"] \ar[d, "E f"'] & A \ar[d, "f"] \\
    E B \ar[r, "\beta"] & B
  \end{tikzcd}
  \qquad \qquad
  f \circ \alpha = \beta \circ E f
\]
It is well-known \cite{freemon} that
the forgetful ``carrier set'' functor of type $\mathbf{Set}^E \rightarrow \mathbf{Set}$
has a left adjoint.
This adjoint maps a set $X$ to the term algebra
$c^E_X : E (E^* X) \rightarrow E^* X$.
Concretely, $c^E_X = [c^m_X]_{m \in E}$ constructs
terms of the form $\underline{m}(t_n)_{n \in \mathsf{ar}(m)}$,
whereas the adjunction's unit $\eta^E_X : X \rightarrow E^* X$
embeds the variables:
\[
  c^E_X \big( \underline{m} \langle t_n \rangle_{n \in \mathsf{ar}(m)} \big) :=
    \underline{m}(t_n)_{n \in \mathsf{ar}(m)}
  \qquad \qquad
  \eta^E_X(x) := \underline{x}
\]
The adjuction's counit
$\epsilon^E_\alpha : \langle E^*A, c^E_A \rangle \rightarrow \langle A, \alpha \rangle$
evaluates terms under their interpretation
$\alpha : EA \rightarrow A$:
\[
  \epsilon^E_\alpha \big( \underline{m}(t_n)_{n \in \mathsf{ar}(m)} \big) :=
    \alpha \big( \underline{m} \langle \epsilon^E_\alpha(t_n) \rangle_{n \in \mathsf{ar}(m)} \big)
  \qquad \qquad
  \epsilon^E_\alpha (\underline{a}) := a
\]
The monad $\langle E^*, \eta^E, \mu^E \rangle$ arising from this adjunction
is called the \emph{free monad} associated with $E$,
and it establishes a connection with the approach
described in \autoref{sec:monadeff}.

The preservation of colimits by left adjoints
means that the initial object in the category $\mathbf{Set}^E$
is given by the algebra
$\mu E = \langle E^* \varnothing, \, c^E_\varnothing \rangle$.
Conversely, $E^* X$ can be characterized as the initial algebra
\[
  [c^E_X, \eta^E_X] \: : \: E(E^*X) + X \: \rightarrow \: E^*X
\]
for a different endofunctor $Y \mapsto E Y + X$.
Given an algebra $[\alpha, \rho] : E A + X \rightarrow A$,
which provides an interpretation
$\alpha^m : A^{\mathsf{ar}(m)} \rightarrow A$ for each function symbol $m \in E$,
and an assignment
$\rho : X \rightarrow A$ of the variables of $X$,
there is a unique algebra homomorphism
$\phi_{\alpha,\rho} :
 \big\langle E^*X, [c^E_X, \eta^E_X] \big\rangle \rightarrow
 \big\langle A, [\alpha, \rho] \big\rangle$:
\[
  \begin{tikzcd}
    E (E^* X) \ar[r, "c^E_X"] \ar[d, "E \phi_{\alpha,\rho}"] &
    E^* X \ar[d, dashed, "!"', "\phi_{\alpha,\rho}"] &
    X \ar[l, "\eta^E_X"'] \ar[d, equal] \\
    E A \ar[r, "\alpha"] &
    A &
    X \ar[l, "\rho"']
  \end{tikzcd}
\]
Note that $\phi_{\alpha,\rho} = \epsilon_\alpha \circ E^*\rho$ and
conversely $\epsilon_\alpha = \phi_{\alpha,\mathrm{id}_A}$.

This universal property provides a foundation
for \emph{effect handlers} \cite{eff},
a programming language construction
which allows a computation to be transformed
by reinterpreting its effects and outcome,
generalizing the well-established use of \emph{exception} handlers.
Using a different kind $F^* Y$ of computations as the target set,
a \emph{handler} $h : E^* X \rightarrow F^* Y$
can be specified using the following data:
\begin{itemize}
  \item a mapping $\rho_h : X \rightarrow F^* Y$
    which provides a computation $\rho_h(x) \in F^* Y$
    meant to be executed when the original computation
    concludes with a result $x \in X$;
  \item an interpretation
    $\alpha_h^m : (F^* Y)^{\mathsf{ar}(m)} \rightarrow F^*Y$
    for each $m \in E$
    providing a computation
    $\alpha_h^m(k_n)_{n \in \mathsf{ar}(m)}$
    to be executed when the original computation triggers the effect $m$.
\end{itemize}
Each argument $k_n \in F^*Y$ of $\alpha_h^m$
corresponds to the (recursively transformed)
behavior of the original computation
resumed by the outcome $n \in \mathsf{ar}(m)$.
We are free to use several of these continuations,
each one potentially multiple times,
to assign an interpretation to the effect.
This flexibility allows handlers to express
a great variety of control flow operators
found in modern programming languages.

\subsection{Final coalgebras} \label{sec:finco} 

The free monad $E^*$ over an effect signature $E$
allows us to represent \emph{finite} computations
with side-effects in $E$
but does not account for \emph{infinite} computations.
By considering the coalgebras for $Y \mapsto EY + X$ instead of algebras,
we can construct an alternative monad $E^\infty$ which 
does not exhibit the same limitation.
Coalgebras are ubiquitous in computer science,
where they appear in the guise of automata and transition systems.
Their use in the context of algebraic effects
therefore presents the additional advantage of
establishing a connection with the associated \emph{operational}
style of semantics.

Concretely, a coalgebra for the endofunctor $Y \mapsto EY + X$
equips a set of states $Q$ with a transition function
$\delta : Q \rightarrow E Q + X$
describing what happens when the computation
is in a given state $q \in Q$:
\begin{itemize}
  \item if
    $\delta(q) = \underline{m} \langle q'_n \rangle _{n \in \mathsf{ar}(n)}$,
    the computation triggers the effect $m \in E$,
    and continues in state $q'_n$ when it is resumed
    by the outcome $n \in \mathsf{ar}(m)$;
  \item if $\delta(q) = \underline{x}$,
    the computation terminates with the result $x \in X$.
\end{itemize}
We define $\langle E^\infty X, \, d^E_X \rangle := \nu Y \,.\, E Y + X$
as the final such coalgebra,
which satisfies the universal property:
\[
  \begin{tikzcd}
    Q \ar[r, "\delta"] \ar[d, dashed, "!"', "\psi_\delta"] &
    E Q + X \ar[d, "E \psi_\delta + \mathrm{id}_X"] \\
    E^\infty X \ar[r, "d^E_X"] &
    E (E^\infty X) + X
  \end{tikzcd}
\]
An applicable construction of terminal coalgebras
can be found in \cite{koind}.


Like $E^*$,
the terminal coalgebra construction $E^\infty$ forms a monad
\cite{coalgmon},
but it can express infinite as well as finite computations
with effects in $E$.
This is used to great effect
in \emph{interaction trees} \cite{itrees},
a data structure designed along these lines
and formalized using coinductive types
in the Coq proof assistant.
A~comprehensive library
provides proof principles and categorical combinators
for interactions trees,
and they are used in the context of the DeepSpec project \cite{deepspec}
to interface disparate certified components.

Nevertheless,
there are limitations to this approach.
In particular,
infinite computations often exhibit \emph{silently divergent} behaviors
(infinite loops).
Modeling these behaviors requires the introduction of a null effect $\tau : \mathbbm{1}$
in the signature $E$,
which coalgebras can then use to delay any interaction.
The elements of $E^\infty X$ must then be considered \emph{up to $\tau$}
(that is, in the context of an algebraic theory
which includes the equation $\tau(x) = x$).
This requires the use of sophisticated simulation techniques
to take into account the distinction between finite iterations
$(\tau^*)$ and silent divergence ($\tau^\omega$).

Less constructively,
we can model silent divergence as its own effect $\bot : \varnothing$.
We will see in the next section that
game semantics can be read as a principled treatment of
this approach,
which reestablishes a connection with algebras
and denotational semantics.



\section{Strategies for uninterpreted effects} \label{sec:effstrat} 


The theory of algebraic effects has a limited scope:
it is intended to be used in conjunction with
existing approaches to programming language semantics
to facilitate the treatment of computational side-effects.
By contrast,
game semantics is its own approach to denotational semantics.
Game models often feature rich, high-order compositional structures.
This reflects the languages they are designed to interpret
and the origins of the technique
in the semantics of linear logic.
On the other hand,
the principles underlying the \emph{construction}
of game models
are somewhat more hazy,
and a huge variety of approaches have been proposed.

Nevertheless,
I~begin this section by attempting to give a high-level account
of what could be dubbed the \emph{classical} approach,
in line with \cite{gsll,gsllaj,pcfajm,pcfho}.
By reading algebraic signatures
as simple games,
I~then deploy some of the techniques used in game semantics
to construct a particularly pleasant model of algebraic effects.
This model can be characterized by specializing
the theory of algebraic effects
to the category $\mathbf{DCPO}_{\bot!}$
of directed-complete pointed partial orders
and strict Scott-continuous functions \cite{domth}.
Notably, the reconciliation operated by game semantics between
denotational and operational semantics
finds a formal expression in the coincidence
between the initial algebras and terminal coalgebras
of endofunctors in $\mathbf{DCPO}_{\bot!}$.


\subsection{Games and strategies} 

\newcommand{\pro}{\mathsf{P}}
\newcommand{\opp}{\mathsf{O}}

The games used in game semantics involve two players:
the \emph{proponent} $\pro$ and the \emph{opponent} $\opp$.
The player $\pro$ represents the \emph{system} being modeled,
while $\opp$ represents its \emph{environment}.
The games we will consider are sequential and alternating:
the opponent opens the game by playing first,
after which the two players contribute every other move.
Game semantics sometimes use simple notions of payoffs,
in which case only the \emph{winning} strategies will be considered,
but by contrast with game theory the focus is primarily
on the \emph{structure of the interaction} between the two players
rather than rational behavior and notions of equilibria.
The games I will consider here
do not use any kind of winning condition.

Traditionally,
a game $G$ is specified by a set of \emph{moves}
$M_G = M_G^\opp \uplus M_G^\pro$
partitioned into opponent and proponent moves.
Then the \emph{plays} of the game $G$
are finite sequences of the form
$m_1 \, \underline{m}_2 \, m_3 \, \underline{m}_4 \cdots$,
where $m_1, m_3, \ldots \in M^\opp$
are opponent moves
and $\underline{m}_2, \underline{m}_4, \ldots \in M^\pro$
are proponent moves.
The set $P_G$ of valid plays of $G$
is often restricted further,
to account for the additional structure
of the particular game model at hand.
In any case,
the objects of interest are then
the \emph{strategies} for $\pro$,
which can be modeled as prefix-closed sets of plays
$\sigma \subseteq P_G$
which prescribe at most one proponent action
in any particular situation:
\begin{equation} \label{eqn:determinism}
  \forall \, s \in P_G^\mathrm{odd} \mathrel\ldot
  \forall \, \underline{m}, \underline{m}' \in M^\pro \mathrel\ldot
  s \, \underline{m}, \, s \, \underline{m}' \in \sigma
  \: \Rightarrow \:
  \underline{m} = \underline{m}'
  \,.
\end{equation}
Although plays are finite,
infinite behaviors can be modeled as
prefix-closed sets of finite approximations.

Categories of games and strategies
can then be constructed.
The objects are games.
The morphisms are strategies $\sigma : A \rightarrow B$
which play a combination of
the game $A$ as the opponent $\opp$ and
the game $B$ as the proponent $\pro$,
starting with an opening move
from the environment in $B$.
Game semantics is related to linear logic \cite{gsll},
and categories of games and strategies often come with a rich structure,
for example:
\begin{itemize}
\item
the game $A \with B$ is played as $A$ or $B$
at the discretion of the opponent,
\item
in the game $A \otimes B$, the games $A$ and $B$ are played side by side,
\item
the game $!A$ allows multiple copies of $A$ to be played
at the discretion of the opponent, and
\item
the game $A^\bot$ reverses the roles of $\opp$ and $\pro$.
\end{itemize}

There are infinite variations on this basic setup,
which have been used to model
imperative programming \cite{gsia},
references \cite{gsgr},
advanced control structures \cite{gscontrol},
nondeterminism \cite{gsfnd,gscnd,gsndsheaves,gseia,nacgs,ifcastrat,rbgs-cal},
concurrency \cite{asfgc}, etc.
Another line of research
explores more fundamental variations on constructions
of game and strategies
\cite{cgames,agames,agames2,tgames},
attempting to provide simpler models of advanced features
and to ground game semantics in a more systematic approach.


\subsection{Strategies for effect signatures} 


Effect signatures can be read as particularly simple games \cite{ifcastrat,rbgs-cal}.
Under this interpretation,
a computation represented as a term in $E^* X$
proceeds in the following way:
\begin{itemize}
  \item the computation chooses a function symbol $m \in E$,
  \item the environment chooses an argument position $n \in \mathsf{ar}(m)$.
\end{itemize}
This process is iterated
until eventually
the computation chooses a variable $x \in X$ rather than a function symbol,
terminating the interaction.
In other words,
a term $t \in E^* X$
can be interpreted as a strategy
for a simple game derived from $E$ and $X$.
We can exploit this analogy to build a model of
computations with side-effects
which mimics the construction of strategies in game semantics.

\begin{definition}[Costrategies over effect signatures] \label{def:costrat} 
The \emph{coplays} over an effect signature $E$
with results in a set $X$
are generated by the grammar:
\[
  s \in \bar{P}_E(X) ::=
    \underline{x} \mid
    \underline{m} \mid
    \underline{m} n s
  \qquad
  (x \in X, \: m \in E, \: n \in \mathsf{ar}(m))
\]
The set $\bar{P}_E(X)$ is ordered by a \emph{prefix} relation
${\sqsubseteq} \: \subseteq \: \bar{P}_E(A) \times \bar{P}_E(A)$,
which is the smallest relation satisfying:
\[
  \underline{x} \sqsubseteq \underline{x}
  \qquad
  \qquad
  \underline{m} \sqsubseteq \underline{m}
  \qquad
  \qquad
  \underline{m} \sqsubseteq \underline{m} n t
  \qquad
  \qquad
  s \sqsubseteq t \: \Rightarrow \: \underline{m} n s \sqsubseteq \underline{m} n t
\]
In addition, the \emph{coherence} relation
${\coh} \subseteq \bar{P}_E(X) \times \bar{P}_E(X)$
is the smallest relation satisfying:
\[
  \underline{x} \coh \underline{x}
  \qquad
  \qquad
  \underline{m} \coh \underline{m}
  \qquad
  \qquad
  \underline{m} \coh \underline{m}ns
  \qquad
  \qquad
  (n_1 = n_2 \Rightarrow s_1 \coh s_2)
  \: \Rightarrow \:
  \underline{m} \, n_1 s_1 \coh \underline{m} \, n_2 s_2
\]
Then a \emph{costrategy}
over the effect signature $E$ with results in $X$
is a downward-closed set $\sigma \subseteq \bar{P}_E(X)$
of pairwise coherent coplays.
I will write $\bar{S}_E(X)$
for the set of such costrategies.
\end{definition}

Note that by contrast with the usual convention,
the first move is played by the system rather than the environment,
hence my use of the terminology \emph{coplays} and \emph{costrategies}.
Moreover,
formulating the condition (\ref{eqn:determinism})
by using a coherence relation is slightly non-traditional
though not without precedent \cite{cohgs}.
Apart from these details,
Definition~\ref{def:costrat}
is fairly typical of the game semantics approach.

Switching back to the algebraic point of view,
we can interpret the terms of $E^* X$ in $\bar{S}_E(X)$
by defining an algebra
$[\alpha, \rho] : E \bar{S}_E(X) + X \rightarrow \bar{S}_E(X)$
as follows:
\[
  \alpha \big(
    \underline{m} \langle \sigma_n \rangle_{n \in \mathsf{ar}(m)}
  \big) :=
    \{ \underline{m} \} \cup
    \{ \underline{m} n s \mid n \in \mathsf{ar}(m), s \in \sigma_n \}
  \qquad \qquad
  \rho(x) := \{ \underline{x} \}
\]
The resulting homomorphism
$\phi_{\alpha,\rho} : E^* X \rightarrow \bar{S}_E(X)$
is an embedding.
However, $\bar{S}_E$ contains many more behaviors, 
including the undefined or divergent behavior $\varnothing$
as well as infinite behaviors,
represented as their sets of finite prefixes.
%
In fact,

\begin{proposition}
$\langle \bar{S}_E , {\subseteq} \rangle$
is a pointed directed-complete partial order.
\begin{proof}
The empty set is trivially a costrategy.
For a directed set $D$ 
of costrategies,
their union $\bigcup D$ is again a costrategy.
Indeed, since $D$ is directed,
any two plays $s_1 \in \sigma_1 \in D$ and $s_2 \in \sigma_2 \in D$
must be coherent:
there exists a strategy $\sigma' \in D$
which includes both $\sigma_1$ and $\sigma_2$,
hence contains both $s_1$ and $s_2$.
\end{proof}
\end{proposition}


This invites us to
give a characterization of $\bar{S}_E(X)$
similar to that of $E^* X$,
by working in the category $\mathbf{DCPO}_{\bot!}$
of pointed dcpos and strict Scott-continuous functions.


\subsection{Complete partial orders} 

Directed-complete partial orders (dcpo for short)
are fundamental to denotational semantics of programming languages.
Before proceeding further,
I summarize a few relevant properties
of the category
of pointed dcpos and strict Scott-continuous functions.

\begin{definition}
A \emph{directed-complete partial order}
$\langle A, {\sqsubseteq} \rangle$
is a partial order with all directed suprema:
any directed subset $D \subseteq A$
has a least upper bound $\sqcup^\uparrow D \in A$,
where \emph{directed} means that
$D$ is non-empty and that
any two $x, y \in D$ have an upper bound $z \in D$.
A \emph{pointed} dcpo has a least element $\bot$.

A \emph{strict Scott-continuous map}
$f : \langle A, {\sqsubseteq} \rangle
\rightarrow \langle B, {\le} \rangle$
between pointed dcpos
is a function between the underlying sets
which preserves the least element $\bot$
and all directed suprema.
The category of pointed dcpos and strict Scott-continous maps
is named $\mathbf{DCPO}_{\bot!}$.
\end{definition}

The category $\mathbf{DCPO}_{\bot!}$ is complete and cocomplete,
as well as symmetric monoidal closed with respect to the \emph{smash} product.
The cartesian product
$\prod_{i \in I} \langle A_i, \sqsubseteq_i \rangle$
is as expected:
the underlying set $\prod_{i \in I} A_i$ is ordered component-wise
and $(\bot_i)_{i \in I}$ is the least element.
The smash product $A \otimes B$ is obtained
by identifying all tuples of $A \times B$
in which at least one component is $\bot$.
The coproduct $A \oplus B$ is called the \emph{coalesced sum}.
It is similar to the coproduct of sets
but identifies $\iota_1(\bot_A) = \iota_2(\bot_B) = \bot_{A \oplus B}$.

The \emph{lifting} comonad $(-)_\bot$
associated with the adjunction between $\mathbf{DCPO}_{\bot!}$
and $\mathbf{DCPO}$
extends a dcpo with a new least element $\bot$.
It can be used to represent (merely) Scott-continuous maps
as strict Kleisli morphisms
$f : A_\bot \rightarrow B$
in $\mathbf{DCPO}_{\bot!}$.
Conversely,
a strict map out of $A_\bot$
can be specified by its (merely Scott-continuous)
action on the elements of $A$.
I will use the same notation
to describe the flat domain construction
$(-)_\bot : \mathbf{Set} \rightarrow \mathbf{DCPO}_{\bot!}$,
left adjoint to the forgetful functor
from $\mathbf{DCPO}_{\bot!}$ to $\mathbf{Set}$.

One remarkable property enjoyed by $\mathbf{DCPO}_{\bot!}$
(and indeed by all $\mathbf{DCPO}_{\bot!}$-enriched categories
\cite{adamek95}),
is that every enriched endofunctor $F$
has both an initial algebra $c : F \mu F \rightarrow \mu F$
and a terminal coalgebra $d : \nu F \rightarrow F \nu F$.
Furthermore, the two coincide
in the sense that $\mu F = \nu F$ and $c^{-1} = d$.


\subsection{Algebraic characterization of strategies} \label{sec:salg} 

The costrategies
for an effect signature $E$ and a set of outcomes $X$
can be characterized as
\begin{equation} \label{eqn:costrat}
  \bar{S}_E(X) \: \cong \: \mu Y \mathrel\ldot \hat{E} Y \oplus X_\bot \:,
\end{equation}
where $\hat{E}$ is defined as follows.
\begin{definition} \label{def:endocpo}
The endofunctor
$\hat{E} : \mathbf{DCPO}_{\bot!} \rightarrow \mathbf{DCPO}_{\bot!}$
associated with the effect signature $E$
is:
\[
    \hat{E} Y :=
      \bigoplus_m \Big( \prod_n Y \Big)_{\!\!\bot}
    \:.
\]
\end{definition}

Algebraically,
the introduction of $(-)_\bot$ in the definition of $\hat{E} Y$
allows the operations to be non-strict.
When an effect $m \in E$ is interpreted,
the resulting computation may be partially or completely defined
even if the continuation always diverges.
In other words, it may be the case that
$\alpha^m(\bot)_{n \in \mathsf{ar}(m)} \neq \bot$.
In terms of game semantics,
this corresponds to the fact that
all odd-length prefixes of coplays are observed,
as witnessed by the case $\underline{m} \in \bar{P}_E$
in Definition~\ref{def:costrat}.

\begin{theorem} \label{thm:stratspec}
For an effect signature $E$ and a set $X$,
the pointed dcpo $\bar{S}_E(X)$
carries the coinciding initial algebra and terminal coalgebra
for the endofunctor $Y \mapsto \hat{E}Y \oplus X_\bot$
on $\mathbf{DCPO}_{\bot!}$.
\begin{proof}
The algebra
$[\hat{c}^E_X, \hat{\eta}] : \hat{E} \, \bar{S}_E(X) \oplus X_\bot
 \rightarrow \bar{S}_E(X)$
can be defined as:
\[
  \hat{c}^E_X \big( \underline{m} \langle \sigma_n \rangle_{n \in \mathsf{ar}(m)} \big) := 
    \{ \underline{m} \} \cup
    \{ \underline{m} n s \mid n \in \mathsf{ar}(m), s \in \sigma_n \}
  \qquad
  \hat{\eta}^E_X(x) :=
    \{ \underline{x} \}
\]
It is easy to verify that
the coplays in $\hat{c}^E_X\big(\underline{m} \langle \sigma_n \rangle_{n \in \mathsf{ar}(m)} \big)$
and $\hat{\eta}^E_X(x)$
are downward closed and pairwise coherent
if the $\sigma_i$'s are.
The coalgebra
$\hat{d}^E_X : \bar{S}_E(X) \rightarrow \hat{E} \bar{S}_E(X) \oplus X_\bot$
can be defined as:
\[
  \hat{d}^E_X(\sigma) :=
  \begin{cases}
    \underline{m}\langle\{ s \mid mns \in \sigma \}\rangle_{n \in \mathsf{ar}(m)}
    & \text{if } \underline{m} \in \sigma \\
    \underline{x}
    & \text{if } \underline{x} \in \sigma \\
    \bot
    & \text{otherwise}
  \end{cases}
\]
The coherence condition on $\sigma$
ensures that the cases are mutually exclusive
and that $\hat{c}^E_X$ and $\hat{d}^E_X$ are mutual inverses.
Thanks to the coincidence of initial algebras and terminal coalgebras
in $\mathbf{DCPO}_{\bot!}$,
this is enough to establish the initiality of
$\langle \bar{S}_E(X), [\hat{c}^E_X, \hat{\eta}^E_X] \rangle$
and the terminality of
$\langle \bar{S}_E(X), \hat{d}^E_X \rangle$.
\end{proof}
\end{theorem}

While much more general constructions
of free algebras in dcpos have been described \cite{dcpoalg},
they tend to be complex.
At the cost of a restriction to effect signatures and \emph{sets} of variables,
costrategies provide a simple construction
with a transparent operational reading.
It may also be possible to extend this construction
to incorporate limited forms of equational theories
by acting on the ordering of coplays.




\subsection{Strategies as ideal completions} \label{sec:sic} 

The algebraic characterization of costrategies given above
invites us to consider more closely the relationship between
$E^* : \mathbf{Set} \rightarrow \mathbf{Set}$
and
$\bar{S}_E : \mathbf{Set} \rightarrow \mathbf{DCPO}_{\bot!}$.
It turns out the costrategies in $\bar{S}_E(X)$
can be constructed as the ideal completion of $E^*(X_\bot)$.

\begin{definition}
An \emph{ideal} of a partial order $A$
is a downward closed directed subset of $A$.
I will write $\mathcal{I}A$
for the set of ideals of $A$,
ordered under set inclusion.
\end{definition}

The ideals of $A$ form a dcpo;
if $A$ has a least element,
then $\mathcal{I}A$ is pointed dcpo.
In fact, $\mathcal{I}A$ is the \emph{free} dcpo generated by
the partially ordered set $A$,
as expressed by the adjunctions:
\[
  \begin{tikzcd}[sep=small]
    \mathbf{DCPO}
      \ar[rr, shift right=2, "U"']
      \ar[dd, shift right=2, "(-)_\bot"']
    & \text{\scriptsize $\bot$} &
    \mathbf{Pos}
      \ar[ll, shift right=2, "\mathcal{I}"']
      \ar[dd, shift right=2, "(-)_\bot"']
    \\
    \text{\scriptsize $\dashv$} & &
    \text{\scriptsize $\dashv$}
    \\
    \mathbf{DCPO}_{\bot!}
      \ar[rr, shift right=2, "U"']
      \ar[uu, shift right=2, "U"']
&
    \text{\scriptsize $\bot$} &
    \mathbf{Pos}_\bot
      \ar[ll, shift right=2, "\mathcal{I}"']
      \ar[uu, shift right=2, "U"']
  \end{tikzcd}
\]
The unit ${\downarrow} : A \rightarrow \mathcal{I}A$
embeds a partial order $A$ into its completion.
A (strict) Scott-continuous map
of type
$f : \mathcal{I}A \rightarrow B$
can be specified by its (strict) monotonic action
$f({\downarrow} a)$
on the elements $a \in A$.


\begin{definition}[Ordering terms]
For an effect signature $E$ and
a partial order $\langle X, {\le} \rangle$,
we extend $E X$ to a partial order
$E \langle X, {\le} \rangle := \langle E X, {\sqsubseteq} \rangle$
by defining $\sqsubseteq$ using the rule:
\[
  \begin{prooftree}
    \hypo{
      \forall n \in \mathsf{ar}(m) \mathrel\ldot x_n \le y_n
    }
    \infer1{
      \underline{m} \langle x_n \rangle_{n \in \mathsf{ar}(m)} \sqsubseteq
      \underline{m} \langle y_n \rangle_{n \in \mathsf{ar}(m)}
    }
  \end{prooftree}
\]
If $\langle X, {\le} \rangle$
has a least element $\bot$,
we extend $E^* X$ to a partial order
$\langle E^*X, {\sqsubseteq} \rangle$
using the inductive rules:
\[
  \begin{prooftree}
    \hypo{
      \forall n \in \mathsf{ar}(m) \mathrel\ldot t_n \sqsubseteq t_n'
    }
    \infer1{
      \underline{m}(t_n)_{n \in \mathsf{ar}(m)} \sqsubseteq
      \underline{m}(t_n')_{n \in \mathsf{ar}(m)}
    }
  \end{prooftree}
  \qquad \qquad
  \begin{prooftree}
    \hypo{x \le y}
    \infer1{\underline{x} \sqsubseteq \underline{y}}
  \end{prooftree}
  \qquad \qquad
  \begin{prooftree}
    \infer0{\underline{\bot} \sqsubseteq t}
  \end{prooftree}
\]
\end{definition}

Here
the elements of $E^*X$ are interpreted as \emph{partial} terms,
where the special variable $\bot \in X$
indicates a lack of information about a given subterm.
In particular, consider
a situation where $t_1, t_2 \sqsubseteq t$.
This means that $t_1$ and $t_2$ are both truncated versions
of the more defined term $t$
and are therefore compatible in the following sense:
although $t_1$ may be defined where $t_2$ is not and vice versa,
they will not conflict on any of their defined subterms
and they can be merged into $t_1 \sqcup t_2$.
By using the ideal completion,
we can extend this procedure to arbitrary directed sets,
enabling the construction of infinite terms.


%
%

\begin{theorem} \label{thm:idealstrat}
For an effect signature $E$ and a set $X$,
the following partial orders are isomorphic:
\[
    \bar{S}_E(X) \: \cong \: \mathcal{I} E^* (X_\bot)
\]
\begin{proof}
It suffices to show that $\mathcal{I} E^*(X_\bot)$
satisfies the characterization of $\bar{S}_E(X)$
given by Theorem~\ref{thm:stratspec}.
We can proceed in the same way.
The algebra
$[\hat{c}^E_X, \hat{\eta}^E_X] :
 \hat{E}\mathcal{I} E^*(X_\bot) \oplus X_\bot \rightarrow
 \mathcal{I} E^*(X_\bot)$
is defined by:
\[
  \hat{c}^E_X \big(
    \underline{m} \langle {\downarrow} \, t_n \rangle_{n \in \mathsf{ar}(m)}
  \big) :=
    {\downarrow} \, \underline{m}(t_n)_{n \in \mathsf{ar}(m)}
  \qquad
  \hat{\eta}^E_X(x) := {\downarrow}\, \underline{x}
\]
The coalgebra
$\hat{d}^E_X :
 \mathcal{I} E^*(X_\bot) \rightarrow
 \hat{E} \mathcal{I} E^*(X_\bot) \oplus X_\bot$
can be defined as:
\[
  \hat{d}^E_X \big( {\downarrow}\, \underline{m}(t_n)_{n \in \mathsf{ar}(m)} \big) :=
    \underline{m} \langle {\downarrow} t_n \rangle_{n \in \mathsf{ar}(m)}
  \qquad
  \hat{d}^E_X({\downarrow}\, \underline{x}) :=
    \underline{x}
\]
As before,
it is easy to check that the required conditions hold
and that $\hat{c}^E_X$ and $\hat{d}^E_X$ are mutual inverses.
\end{proof}
\end{theorem}

Finally,
it has been shown \cite{coalgideal} that
$\mathcal{I} E^*(X_\bot) \cong E^\infty(X_\bot)$.
Hence,
Theorem~\ref{thm:idealstrat}
also establishes a connection between
strategies
and the coalgebraic approach
discussed in \autoref{sec:finco}.



\section{Towards algebraic game semantics} \label{sec:multisort} 

The constructions given in the previous section
provide a model of algebraic effects
grounded in an interpretation of effect signatures as games.
Conversely,
while effect signatures are a very restricted class of games,
the analysis above
establishes a blueprint for a broader
reading of games and strategy constructions
under the lens of categorical algebra.

My hope is that by extending the basic framework in various directions,
it would become possible to account for a broad range
of existing game models
and provide insight into the details of their construction.
Ideally,
the resulting theory would decouple various aspects of their design
and provide a general construction toolkit for game semantics.
In this section I outline several possible avenues
for future work in this direction,
and point to some existing work relevant to that goal.

\subsection{Varying the order completion} 

The reformulation of costrategies as ideal completions
presented in \autoref{sec:sic}
invites generalization in at least two different ways.

First,
since it is built from terms rather than plays,
it provides a better starting point
for incorporating equational (and inequational) theories
to the model,
specifically by quotienting
the set of terms $E^*(X_\bot)$
before the ideal completion is applied.
This would permit the use of a broader range of theories
than can be expressed using plays alone
(as suggested at the end of \autoref{sec:salg}).

Moreover,
the ideal completion itself
could be replaced by different order completions.
This could yield the kind of
nondeterministic models
proposed in Bowler et al.~\cite{ifcastrat} or
the \emph{interaction specification} monad of
Koenig \& Shao \cite{rbgs-cal}.
Importantly,
\begin{center}
the \emph{plays}
used in game semantics
correspond to
the \emph{join-irreducible strategies}.
\end{center}
This suggests that
it may be possible to reconstruct the models mentioned above
using variations on the \emph{Bruns-Lakser completion},
presented for example in Bazerman and Puzio \cite[\S3]{bazbl}.
This may require a better understanding
of the ways initial algebra and terminal coalgebra constructions
propagate \cite[\S2.5]{adjtrans} through the adjunctions
defined by order completions---%
this may also shed light on
the relationship between the endofunctors $\hat{E}$ and $E$.

Ultimately,
the goal would be to decouple the
order-theoretic character of a given model
(and the corresponding support of
partial definition, infinite behaviors and nondeterminism),
determined by the order completion used,
from the structure of the interaction which the model captures,
determined by the partially ordered set of terms to be completed.


\subsection{Operations on signatures} 

The endofunctors associated with
effect signatures per Definition~\ref{def:sigendo}
correspond exactly to the
single-variable \emph{polynomial} endofunctors over
the category $\mathbf{Set}$.
Polynomial functors in general
have been studied extensively.
Single-variable polynomials on $\mathbf{Set}$ in particular
have recently been a source of great interest
in modeling dynamical systems \cite{poly},
owing to their remarkable properties.

Specifically,
the category $\mathbf{Poly}$
of single-variable polynomials on $\mathbf{Set}$
has all products ($\times$) and coproducts ($+$),
a separate notion of tensor product ($\otimes$),
with endofunctor composition ($\circ$)
defining yet another monoidal structure.
$\mathbf{Poly}$ is closed with respect to \emph{both}
the cartesian and tensor products,
and even \emph{co-closed} with respect to the tensor product.
It is worth noting as well that
the free monad construction $E^*$
described in \autoref{sec:initalg}
can be internalized as a free monoid monad in $\mathbf{Poly}$.

Many of these structures also appear in the context of
game semantics and linear logic.
In $\mathbf{Poly}$, they rely in large part on
complete distributivity of the underlying category $\mathbf{Set}$;
as such
it is unclear how they carry over to the context of
$\mathbf{DCPO}_{\bot!}$ and Definition~\ref{def:endocpo}.
But this suggests that despite the apparent simplicity
of effect signatures as games,
their expressive power can be significant.


\subsection{Multi-sorted signatures} 

Sequential alternating games
of the kind used in game semantics
can be described using bipartite directed multigraphs \cite{tsg}:
vertices correspond to the possible phases or states of the game,
and are partitioned into opponent and proponent states;
edges out of a particular state correspond to
the possible next moves,
and the plays are paths through the graph.

In the case of effect signatures,
there is only one proponent vertex.
As such, the corresponding games are almost stateless:
every time the proponent $\mathsf{P}$ is back in control,
the game must be reiterated without any change to the rules.
This is due to the \emph{single-sorted} nature of effect signatures:
while \emph{arities} provide the opponent $\mathsf{O}$
with different sets of moves in different situations,
the single \emph{sort} does not permit the same flexibility for $\mathsf{P}$.

From an algebraic perspective,
lifting this restriction means
generalizing the framework to multi-sorted signatures.
This restores the symmetry between the two players,
as emphasized by the following presentation.

\begin{definition}
  A \emph{multi-sorted effect signature} is a tuple
  $E = \langle
    \bar{Q},
    \bar{M},
    \bar{\delta},
    Q,
    M,
    \delta
  \rangle$.
  The components define:
  \begin{itemize}
    \item a set $\bar{Q}$ of \emph{sorts} and a set $Q$ of \emph{arities};
    \item for every sort $q \in \bar{Q}$
      a set $\bar{M}_q$ of \emph{function symbols} and
      for every $m \in \bar{M}_q$
      an arity $\bar{\delta}_q(m) \in Q$;
    \item for every arity $r \in Q$
      a set $M_r$ of \emph{argument positions} and
      for every $n \in M_r$
      a sort $\delta_r(n) \in \bar{Q}$.
  \end{itemize}
\end{definition}

The sorts (proponent states) and arities (opponent states)
can both be understood as \emph{types},
respectively for operations and argument tuples.
The game alternates between
a proponent choice of operation
and an opponent choice of argument position.

To follow the blueprint laid out in \autoref{sec:effstrat},
we must assign endofunctors to multi-sorted signatures.
These will be polynomial functors of multiple variables,
hence we will be working in categories of
$Q$- or $\bar{Q}$-indexed \emph{tuples}
of sets and functions.
The term algebra can then be constructed using
sets of mutually recursive terms and argument tuples,
and we can use the ideal completion to obtain
a strategy model.
Note that we must also pick an initial sort (for costrategies)
or arity (for strategies).

Again the connection with polynomial functors
allows us to draw from a large body of existing work.
For example,
Hyvernat \cite{hyvernat-diag,hyvernat-ext} gives
a model of linear logic based on polynomials of this form.

\section{Conclusion} 

Although much work remains to be done,
looking at game semantics
through the prism of categorical algebra
offers promising avenues of investigation.
Multi-sorted signatures
constitute a low-level representation for
sequential alternating games.
Characterizing existing forms of game semantics using algebraic tools
could reveal interesting structures,
and suggest general principles
for the design of general-purpose models
capable of accounting for the behaviors
of a wide range of heterogeneous components.


\section*{Acknowledgments} 

This work greatly benefited from
conversations with
Arthur Oliveira Vale,
L\'eo Stefanesco,
Paul-Andr\'e Melli\`es, and
Zhong Shao.
It was supported in part by
\href{http://nsf.gov}{NSF} grants
1521523,
1763399, and
2019285.
I~am also grateful for the feedback
provided by the anonymous ACT reviewers
and volume editor Kohei Kishida,
which significantly improved the quality of
the final paper.


\bibliographystyle{eptcs}
\bibliography{references}

\end{document}